# APPLYING THE APRIORI ALGORITHM FOR INVESTIGATING THE ASSOCIATIONS BETWEEN DEMOGRAPHIC CHARACTERISTICS OF IRANIAN TOP 100 ENTERPRISES AND THE STRUCTURE OF THEIR COMMERCIAL WEBSITES

Ali Azimi[1], Azar Kaffashpour[2]

[1]PhD candidate in Management, Ferdowsi University of Mashhad
[2]PhD in Marketing Management, Faculty member at Ferdowsi University of Mashhad, School of Management

## ABSTRACT

*This study was conducted with the main aim to investigate the relationships between the demographic characteristics of companies and the facilities required for their commercial websites. The research samples were the top 100 Iranian companies as ranked by the Iranian Industrial Management Institute (IMI); the method applied is data-mining, using the Association Rules through the A-priori algorithms. To collect the data, an author-modified checklist has been utilized covering the three areas of the facilities within commercial websites, i.e. fundamental, information–providing, and service-delivering facilities. Having extracted the association rules between the mentioned two sets of variables, 68 rules with a confidence rate of 90% and above were obtained and, based on their significance, classified into two groups of must-have and should-have requirements; a recommended package of facilities is hitherto offered to other companies which intend to enter e-commerce through their commercial websites with regards to each company's unique demographic characteristics.*

## 1. INTRODUCTION

One of the main tools for every business in e-commerce is, by all means, its commercial website. The structure of commercial websites can be studied from different viewpoints. Aspects of content and technique, as well as the interaction facilities are items which may be used to assess the different facilities existing in such websites. The structure of commercial websites and the selection of facilities they utilize for doing business have great significance in their success (or failure) in e-commerce, regarding the fact that the websites of companies are their representative on the Internet as well as their means of communication with customers, competitors, commercial partners, and other business stake-holders [1].

On the other hand, businesses differ with each other in their fields of activities, are not the same regarding their type of ownership, and, of course, vary according to their age (i.e. their life span since their establishment); in other words, the demographic characteristics of a business (including its field of activity, type of ownership, and age) are among the factors potentially affecting other aspects of the business. Therefore, it can be esteemed necessary for business owners to be aware of the requirements of their commercial websites for entering the e-commerce in accordance with the unique characteristics of their company, specifically in such demographic areas as the age, ownership type, and industry group; besides, the providers of infrastructures of electronic commerce (i.e. the governmental policy-makers and web-facilities providers) should evaluate such requirements of each group of businesses in alignment with their characteristics, strategies, and environment if they want to efficiently fulfill those needs [2]. These two above-mentioned sets of factors (demographic characteristics of the company, on one hand, and the structure of its commercial website on the other) are the necessities highlighted in the present research; 20 important items for designing a commercial website have been introduced using an author-designed checklist (Appendix A). Besides, companies which own commercial websites are apt to be categorized with respect to different aspects.

Demographic characteristics of each firm can be used as one possible criterion for categorizing different businesses. In our research, three demographic characteristics were individually investigated for the studied businesses, namely: 1) age (since the firm's establishment), 2) type of ownership (private, semi-private, or governmental), and 3) their industry (products or services).Reports of the demographic situation of each firm with respect to the 3 groups on the evaluation checklist (i.e. fundamental, information-related, and service-delivering facilities) were separately obtained as well; next, in order to determine the relationships between these two sides (the demographic characteristics of the companies and the structure of their commercial websites), a data-mining approach was applied. After studying the similar researches which had used data-mining, and post to conducting a comparison of the description of their research questions as well as their methods and procedures, the A-priori algorithm was selected from among other current algorithms, with the aim to extract the Association Rules that may clarify any possible relationships available.

## 1.1 Theoretical Background

Previous studies on investigating the process of e-commerce through commercial websites of businesses and especially their relation with the structural characteristics of different companies have mainly focused on basic concepts of e-commerce and commercial websites designing. Since data-mining was applied to conduct this research, it will be practical here to review other researches that have used the same method.

Papastathopoulo and Avlonitis [3] have attempted, firstly, to determine several indexes for different businesses in respect of their policies in using the web space and, secondly, to pinpoint the relationships between the type of use of the world-wide web in general with the demographic characteristics of the top companies of Greece, having, finally, concluded that the type of web-use profiles in those companies could be classified into distinct groups.

Zantidis and Nicholas [4] in investigating the types of Internet usage in Greece and its impact on the procurement of e-commerce in that country, consider the advancement of information technology, the development of the required equipment for creating e-commerce websites, and expansion of Internet use as three of the main concerns of Greek government, coming to the conclusion that the *governmental incentives*, the *reduction of the costs of the necessary technologies*, and the *public acceptance of the Internet as the medium for buying and selling* could be regarded as crucial factors for success in e-commerce.

Hacket and Parmano [5] have dealt with the question whether surfing the homepages of the websites suffices for assessing the rate of their accessibility (as one of the criteria in evaluation and ranking of websites) or not; they introduce WAB factor (Web Availability Barriers) and recommend that at least the contents of the main pages of a website need to be assessed if we want to make a sound judgment over its level of availability.

Babagoli and Khantarzadeh [6], in their investigation of the barriers of developing e-commerce in Iran, identify and prioritize the environmental barriers against the e-commerce expansion, thus determining the significance of each observed index, introducing some practical solutions, and finally proposing a conceptual model for optimal administration of e-commerce in the country. Davarpanah and Khaleghi [1] tried to evaluate Iranian websites based on a systemic investigation of their unique qualities; they've found out that private websites in Iran are weaker than governmental ones in respect to presenting updated contents in various languages as well as in providing the related links to other websites and/or internet information sources. Salmani and Nasrollahzadeh [7] proposed a criterion for assessing the e-commerce in Iran through investigating several discussions related to collecting statistical data and applying modern commercial technologies; they, furthermore, reviewed the considerable experiences and achievements of pioneering global organizations in this field, coming up with some suggestions for improving the assessment of e-commerce in developing-countries including Iran.

## 2. MATERIAL AND METHODS

Since this research was carried out using data-mining approach, there was no definite hypothesis beforehand and the existing principles (rules) were obtained post to performing data-mining and without any pre-assumptions on the probable emerging results; of course, the purposes of the present study can be raised as the following questions:

1. Are there any relationships between demographic characteristics of companies (including their age, type of ownership, and industry group) and the facilities and structural characteristics of their commercial websites?

2. In case there are such relationships available, what suggestions and recommendations could be put forward for other companies planning to enter e-commerce and use the facilities of their commercial websites more effectively?

This study aims to investigate the relationships between demographic characteristics of companies and the facilities in their commercial websites. Commercial websites of top 100 Iranian enterprises of 2011 were investigated and publicly announced in December, 2011 by Iran's Industrial Management Institute (IMI). The main criterion for this yearly national ranking is the annual sale rate provided by the financial information of these businesses in the previous year.

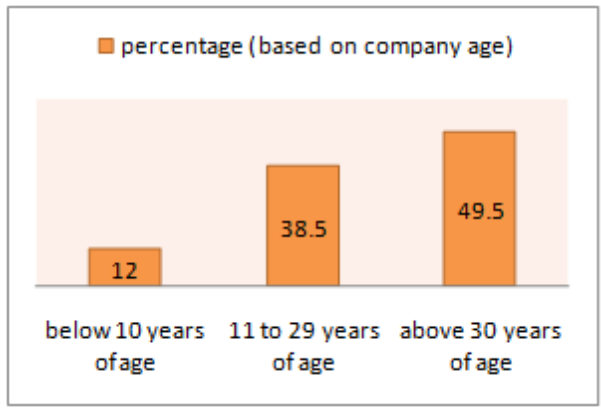

Diagram 1. Distribution of the surveyed enterprises based on the age of the company

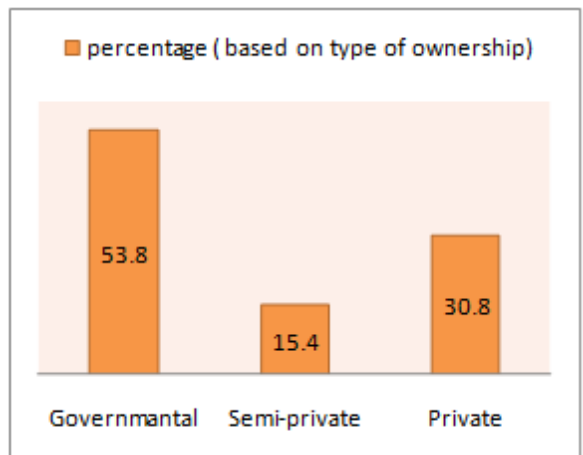

Diagram 2. Distribution of the surveyed enterprises based on the type of ownership of the company

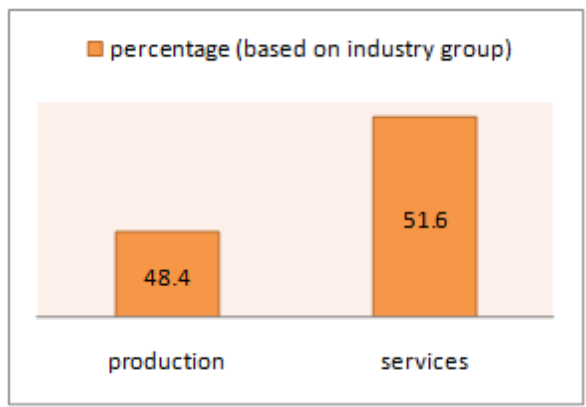

Diagram 3. Distribution of the surveyed enterprises based on their industry groups

Both library and the Internet resources were utilized to conduct studies on e-commerce and the usage of websites for businesses; field approaches in the present study included filling in the checklists to determine the characteristics of the commercial websites of the selected businesses with the aim to gather data for extracting the variables and excavating the rules. After comparing and merging the current methods of website evaluation, a checklist was modified including 20 questions for evaluating the research community in the three categories of basic and technical (labeled here as *fundamental*) facilities of the websites (questions 1-5), service-delivering facilities of the websites (questions 6-10), and information-providing facilities of websites (questions 11-20). This checklist passed a validity test through getting reviewed by 4 members of scientific board of universities who had experiences of similar researches in their fields (the original checklist contained 28 items, 8 of which being omitted through this check).

To test the reliability of the checklist, all these commercial websites were assessed twice with an interval time of 30 days, first by a group of 11 students of different majors of computer engineering and then by one of the authors individually. Cronbach's alpha was obtained as the indicator of reliability index, using SPSS 17 software and the results determined a correlation coefficient of 0.94 for the tools of this research.

As the next step, all the 100 top companies (which were introduced by Industrial Management Institute in 2011) were selected as the research community and therefore no sampling was necessary to take place. Data produced in the first section of the checklist (the demographic characteristics of each company: its age, type of ownership, and industry group) were used as input data in different phases of data-mining. Data of the second section was related to investigating the existence or lack of each of the 20 facilities in the commercial websites of those companies; all data in these two sections were fed into SPSS Clementine 11 Data-mining Software with the aim to excavate the association rules between these two  sets of variables (demographic characteristics of companies and the structural characteristics of their commercial websites); the rate of confidence as a main factor in data-mining was set to 90% and above; the applied data-mining algorithm in the present study was A-priori. A brief look at data-mining approach can be presented as follows:

Data-mining is defined as the process of discovering the hidden knowledge within piles of data or discovering the spectacular, unexpected, and valuable structures  and relations within a great deal of data. In other words, data-mining means analyzing and exploring a great deal of data with an aim to discover meaningful patterns and rules [8]. Identifying and extracting patterns and regularities in massive data repositories has been a focused theme in data mining research for almost over a decade [9].

In defining "Association Rules" and explaining the concepts of rate of "Confidence" and "Support", it can be noted that Association Rule is a type of specific operation for data-mining that will search for each and every possible relationship between multitude characteristics in large data sets. This operation includes studying the characteristics or trends which are interrelated, in order to extract the rules and to quantify the relationships between two or more characteristics. Association Rules are firstly defined as discovered relations and then are accompanied by the two criteria of Support and Confidence as follows:

- Rate of confidence: instances of cases in a data-set in which when the antecedent is present, the consequent will be, too.

- Rate of support: instances of cases in a data-set in which both antecedent and the consequent are present [10].

Association rules is a novel data mining technique that has been mainly used for data description, exploration and prediction in knowledge discovery and decision support systems. The association rule mining algorithm is modified to handle the user-defined input constraints. Associative classification is provided with a large number of rules, from which a set of quality rules are chosen to develop an efficient classifier [11]. Data-miners prefer those rules which have higher degrees of Confidence, of Support, or of both simultaneously. It should be pointed out that stronger Association Rules are those which enjoy higher degrees of Confidence and Support [12].

Regarding the application of data-mining in web space, it is argued that web pages represent various information that is easy and efficient way to meet the user requirement; a large type of data contains in various web page from different website and domain [13]; to delve into the algorithm applied in the present paper, i.e. in Apriority Algorithm, for extracting the association rules the problem of association rule mining is defined as: Let I= {i1, i2, …..in} be a set of n binary attributes called items. Let D= {t1, t2,…., tm} be a set of transactions called the database. Each transaction in D has a unique transaction ID and contains a subset of the items in I. A rule is defined as an implication of the form X=>Y where X, Y C I and X_ Y=ø. The sets of items X and Y are called antecedent (Left-hand side) and consequent (Right-hand side) of the rule. To select interesting rules from the set of all possible rules, constraints on various measures of significance and interest can be used. The best-known constraints are minimum thresholds on support and confidence [14].

A-priori algorithm can be held as one of the most significant findings in the history of excavating the Association Rules. This algorithm has made use of the fact that all the subsets of frequent items are, themselves, frequent. This algorithm, however, will process the database just once and therefore has less calculation capacity in comparison with other algorithms [15].

## 3. RESULTS

As it can be observed in several researches mentioned earlier in the literature review, the developed countries, in comparison with the developing countries including Iran, have moved more consistently and scientifically towards enhancing their e-commerce structures; studies conducted in Iran are still mostly confined to trying to define the concept of the term e-commerce or to justify its importance in today's business world and only recently have some solutions for existing challenges gained an amount of attention.

Table 1. Frequency of the existence of the studied facilities in the commercial websites of the companies

| number | Description of the facility | Total% | Type of ownership | | Industry Group | | Company age | | |
|---|---|---|---|---|---|---|---|---|---|
| | | | Gov.owned % | Private/Semi private owned % | production% | services% | Below 10 yrs% | 11-29 yrs% | Above 30 yrs% |
| 1 | "About Us" page | 95.60 | 97.96 | 92.86 | 97.73 | 93.62 | 100 | 97.14 | 93.33 |
| 2 | "Contact Us" page | 97.80 | 95.92 | 100.00 | 100.00 | 95.74 | 100 | 94.29 | 100.00 |
| 3 | Search Inside the site | 74.73 | 75.51 | 73.81 | 72.73 | 76.60 | 81.82 | 85.71 | 64.44 |
| 4 | English Homepage | 76.92 | 89.80 | 69.05 | 70.45 | 82.98 | 36.36 | 88.57 | 77.78 |
| 5 | Selected contents in English | 74.73 | 81.63 | 66.67 | 70.45 | 78.72 | 27.27 | 88.57 | 75.56 |
| 6 | Related links | 76.92 | 85.71 | 71.43 | 70.45 | 82.98 | 90.91 | 82.86 | 68.89 |
| 7 | Links to news and information | 68.13 | 65.31 | 71.43 | 63.64 | 72.34 | 72.73 | 74.29 | 62.22 |
| 8 | Personnel log-in | 63.74 | 53.06 | 76.19 | 59.09 | 68.09 | 81.82 | 51.43 | 68.89 |
| 9 | Research department | 58.24 | 57.14 | 59.52 | 65.91 | 51.06 | 54.55 | 45.71 | 68.89 |
| 10 | Links to related sites | 70.33 | 69.39 | 71.43 | 65.91 | 74.47 | 72.73 | 80.00 | 62.22 |
| 11 | Site Map | 72.53 | 71.43 | 71.43 | 63.64 | 78.72 | 63.64 | 74.29 | 71.11 |
| 12 | Specific page-title | 61.54 | 55.10 | 69.05 | 61.36 | 61.70 | 63.64 | 60.00 | 62.22 |
| 13 | Reasonable load time | 86.81 | 91.84 | 80.95 | 84.09 | 89.36 | 81.82 | 88.57 | 88.89 |
| 14 | General introduction of its field | 91.21 | 95.92 | 85.71 | 93.18 | 89.36 | 81.82 | 91.43 | 93.33 |
| 15 | Detailed description of the company | 57.14 | 53.06 | 61.90 | 59.09 | 55.32 | 54.55 | 54.29 | 60.00 |
| 16 | Detailed sub-menus | 79.12 | 69.39 | 90.48 | 79.55 | 78.72 | 72.73 | 77.14 | 82.22 |
| 17 | Branches info | 62.64 | 57.14 | 69.05 | 47.73 | 76.60 | 63.64 | 62.86 | 62.22 |
| 18 | Advertises its own products/services | 70.33 | 67.35 | 73.81 | 68.18 | 72.34 | 72.73 | 62.86 | 75.56 |
| 19 | "company at a glance" page | 90.11 | 87.76 | 92.86 | 88.64 | 91.49 | 90.91 | 88.57 | 91.11 |
| 20 | "company at a glance" in English | 68.13 | 71.43 | 64.29 | 59.09 | 76.60 | 54.55 | 77.14 | 64.44 |

As it was indicated in the literature review, tools and contents needed for the present study were obtained from the finding of selected previous researches. Classifying the companies based on their demographic characteristics and their use of web-space is a common trend relevant to this research as well. Investigating the importance of e-commerce and its applications in other countries and considering its barriers in Iran with respect to other successful experiences mentioned in national and international researches is an indicative of the main approach of this research.

Setting the data-mining with a certainty rate of at least 90%, eleven out of the twenty facilities of the commercial websites studied appeared to have a meaningful relationship with the demographic characteristics of their owner companies; 68associationrules (Appendix B) were acquired and classified into two categories: necessary facilities (*must-have*'s), and preferred facilities (*should-have*'s). After investigating the data and excavating the Association Rules, the rate of the existence of each facility in different groups of the studied companies can be sorted based on the highest rate of certainty (100%) to the lowest acceptable rate (90% in this research). The ranking of these 11 facilities of data-mining in the following table (Table 2) is based on the number of the Association Rules excavated, keeping in mind that the A-priori algorithm is programmed to produce Association Rules only within the groups with the pre-determined Confidence rate.

Table 2. Company types ranked by the degree of the availability confidence of the required facilities for their commercial websites

| No. | The facility available in the commercial websites of top 100 Iranian companies | Types of companies (according to their demographic characteristics), sorted by the level of confidence of the mined association rules (ascending) |
|---|---|---|
| **1** | **"Contact-us Page"** | 1. Companies with less than 10 years of age |
| | | 2. Companies with more than 30 years of age |
| | | 3. Companies with private ownership |
| | | 4. Companies in production industries |
| | | 5. Companies with more than 30 years of age AND with private ownership |
| | | 6. Companies with more than 30 years of age AND with governmental ownership |
| | | 7. Companies with 11 to 29 years of age AND in production industries |
| | | 8. Companies with more than 30 years of age AND in production industries |
| | | 9. Companies with more than 30 years of age AND in service-providing industries |
| | | 10. Companies with private ownership AND in service-providing industries |
| | | 11. Companies with governmental ownership AND in production industries |
| | | 12. Companies with governmental ownership |
| | | 13. Companies in service-providing industries |
| | | 14. Companies with 11 to 19 years of age |
| | | 15. Companies with 11 to 29 years of age AND with governmental ownership |
| | | 16. Companies with 11 to 29 years of age AND in service-providing industries |

| 2 | "About-us Page" | 1. Companies with less than 10 years of age |
|---|---|---|
| | | 2. Companies with semi-private ownership |
| | | 3. Companies with 11 to 29 years of age AND with governmental ownership |
| | | 4. Companies with 11 to 29 years of age AND in production industries |
| | | 5. Companies with governmental ownership AND in service-providing industries |
| | | 6. Companies with governmental ownership |
| | | 7. Companies in production industries |
| | | 8. Companies with 11 to 29 years of age |
| | | 9. Companies with governmental ownership AND in production industries |
| | | 10. Companies with more than 30 years of age AND with governmental ownership |
| | | 11 Companies with more than 30 years of age AND in production industries |
| | | 12. Companies with 11 to 29 years of age AND in service-providing industries |
| | | 13. Companies in service-providing industries |
| | | 14. Companies with more than 30 years of age |
| 3 | "general introduction of its field" | 1. Companies with 11 to 29 years of age AND with governmental ownership |
| | | 2. Companies with governmental ownership AND in service-providing industries |
| | | 3. Companies with governmental ownership |
| | | 4. Companies with more than 30 years of age AND in service-providing industries |
| | | 5. Companies with more than 30 years of age |
| | | 6. Companies with 11 to 29 years of age in production industries |
| | | 7. Companies with governmental ownership AND in production industries |
| | | 8. Companies in production industries |
| | | 9. Companies with more than 30 years of age AND with governmental ownership |
| | | 10. Companies with more than 30 years of age AND in production industries |
| | | 11. Companies with 11 to 29 years of age |
| | | 12. Companies with more than 30 years of age AND with private ownership |
| | | 13. Companies with 11 to 29 years of age AND in service-providing industries |
| 4 | **"a brief introduction of the company"** | 1. Companies with semi-private ownership |
| | | 2. Companies with more than 30 years of age AND in production industries |
| | | 3. Companies with 11 to 29 years of age AND in service-providing industries |
| | | 4. Companies with more than 30 years of age AND with governmental ownership |
| | | 5. Companies in service-providing industries |
| | | 6. Companies with more than 30 years of age |
| | | 7. Companies with less than 10 years of age |
| 5 | **the uploading speed of the main pages** | 1. Companies with governmental ownership AND in service-providing industries |
| | | 2. Companies with more than 30 years of age AND with governmental ownership |
| | | 3. Companies with governmental ownership |
| | | 4. Companies with 11 to 29 years of age AND in service-providing industries |
| 6 | **"English homepage"** | 1. Companies with more than 30 years of age AND in service-providing industries |
| | | 2. Companies with governmental ownership AND in service-providing industries |
| | | 3. Companies with 11 to 29 years of age AND with governmental ownership |
| | | 4. Companies with 11 to 29 years of age AND in service-providing industries |
| 7 | **"selected contents in English"** | 1. Companies with governmental ownership AND in service-providing industries |
| | | 2. Companies with 11 to 29 years of age AND with governmental ownership |
| | | 3. Companies with 11 to 29 years of age AND in service-providing industries |
| 8 | **"Related Links"** | 1. Companies with semi-private ownership |
| | | 2. Companies with less than 10 years of age |
| | | 3. Companies with 11 to 29 years of age AND in service-providing industries |
| 9 | **"detailed sub-menus" in their commercial websites** | 1. Companies with semi-private ownership |
| | | 2. Companies with more than 30 years of age AND with private ownership |
| 10 | **"Research Department"** | 1. Companies with semi-private ownership |
| 11 | **specific page-titles** | 1. Companies with more than 30 years of age AND with private ownership |

As it can be observed in Table 2-4, among the total 11 facilities which showed a meaningful relationship with the demographic characteristics of these companies based on the Association Rules obtained from data-mining, the first 4 ones (including the facilities of "Contact-us page", "About-us page", "general introduction of its field", and "a brief introduction of the company") existed in almost all of the commercial websites of the top Iranian companies (in 13, 14 and 7 groups, respectively); it can also be mentioned that these 4 facilities were common among all the companies irrespective of their differences in demographic characteristics. Comparing these facilities with the applied checklist could be conducted as follows:

a) Technical facilities of the website: none
b) Information-providing facilities: 3 items (having "about us" page, "general introduction of its field", and "a brief introduction of the company")
c) Service-delivering facilities: 1 item (having "contact us" page)

The second 7 items (including the facilities of "English homepage", "reasonable load-time", "selected contents in English", "related links", " detailed sub-menus for easier access to different parts of websites", "research department" and " specific page titles") existed only in under­lined some websites and it is clear that the demographic characteristics of *company age*, the *type of ownership*, and the *industry group* can have a significant impact on the necessity of existence or absence of these facilities in the commercial websites of those companies.

This group of facilities is comparable (with the applied checklist) providing that the three sets of facilities existing in the checklist be considered as follows:

a) Technical facilities: 3 items (having "reasonable load-time", "detailed sub-menus for easier access to different parts of the website", and "specific page titles")
b) Information-providing facilities: 3 items (having "English homepage", "selected contents in English", and "research department")
c) Service-delivering facilities: 1 item (having "related links")

## 4. DISCUSSION

Taking the analyses of this research into account, it can be discussed that there are meaningful relationships between some demographic characteristics of the businesses and some of the facilities and structural characteristics of their websites. Of the 20 facilities studied in this paper (having been categorized into the three groups of fundamental, information-providing, and service-delivering facilities), 11 had a strong relationship (based on the rate of confidence of the rules higher than 90%) with the demographic characteristics of their owner companies (i.e. the company age, type of ownership, and industry group).Therefore, businesses that use their commercial websites for doing e-commerce can use the findings of this study if they are going to decide what facilities are required for their commercial websites in accordance to their own demographic characteristics.

Rating the facilities existing in the commercial websites of different groups of companies is based on confidence rate obtained through data-mining (34 association rules with the rate of 95 to 100% have top priority [*must-have*'s] and those with a certainty rate of 90 to 94.9% have a secondary priority [*should-have*'s]).Of course, the facilities proposed in these packages are considered as the minimal tools needed for a business to enter e-commerce by following (or learning from) successful businesses with similar characteristics. Nevertheless, it does not mean that all the required facilities are mentioned here.

Findings of this study are in accordance with the previous researches concluding that Iranian companies, to conduct e-commerce, still face many shortcomings regarding the structure of their commercial websites ([8], [16], and [17]). In clarifying the shortcomings mentioned in previous researches, 3 of the technology-related facilities out of 5 mentioned here had a significant relationship with the demographic characteristics of their companies and out of the 5 items related to service-delivering facilities of the websites, only 2 had a significant relationship with the characteristics of the studied companies, one being held as a necessity (must-have) for all websites while the second as a preferred (should-have) facility for companies with specific demographic characteristics. Finally, 6 out of the 10 questions related to the information-providing facilities of the websites had a significant relationship with the demographic characteristics of their owner company, which, by itself, means that the commercial websites of Iranian top companies are mainly focused on information-providing; these latter findings of the present study, too, are in agreement with the results of the previous researches ([18], [19], [20], and [21]).

# 5. CONCLUSIONS

To conclude the present paper in a more practical order, the authors categorized the *Necessary* and the *preferred* facilities in the following manner:

    a) Necessary (Must-have):

        1- "Contact Us" facility
        2- "About Us" page
        3-"General introduction of its field"
        4- "A brief introduction of the company"

These findings which complement the previous researches ([20] and [21]) show that the majority of these commercial websites have solely focused on providing general information and this way of using the web-space facilities is far too limited for commercial websites of top companies at national level.

    b) Preferred (Should-have):

        1-"English homepage"
        2-"Selected contents in English"
        3-"Research department" section

The first 2 items are in accordance with the results of the previous studies ([1] and [3]); in other words, willingness of the Iranian companies to introduce themselves in English is an indicative of their caring for providing the infrastructure which several studies have mentioned as an essential prerequisite for entering global markets ([21], [1], and [17]).The third facility regarding the presentation of the company's researches on the Internet is *reassuring* for these companies themselves, *promising* for the policy-makers of e-commerce, and *instructive* for other Iranian enterprises that may intend to learn from top Iranian companies on how to enter the e-commerce through their commercial websites.

This research, besides its achievements, had the following limitations:

1.Some websites of the studied companies (9 out of 100) were not accessible in our two assessing trials with the interval of 30 days and therefore could not be included in this research.

2.Only the company age, the type of ownership, and industry group were investigated here from among various demographic characteristics; variables like the number of employees in the company, the employees' sex distribution, or the CEO's gender were not taken into account in our study.

3. Companies studied in this research were held as top 100 companies according to the latest public rating list of the Industrial Management Institute. This may strengthen the validity and pattern-providing value of our findings, but cannot, in some aspects, be applicable for certain groups of businesses like small and medium-sized enterprises (SMEs) or start-up ventures, due to their incomparable age and financial background with those of the top companies.

Considering the significance of structural characteristics and the existing facilities of the commercial websites of companies as a fundamental requirement for entering the e-commerce and tailoring these different characteristics and subsequently different needs of various businesses, it is suggested that researchers of this field take into account the following issues:

1-future researches can study other groups of companies (in respect of working record (age), size of the company, financial situation, etc.).

2-in separate researches, the demographic characteristics could be replaced (or completed) with other factors like the number of personnel, the CEO's gender, or the distribution of the employees' sex, etc.

3-structural characteristics and facilities of the commercial websites can include other items including the e-shopping tools, electronic payment, or providing options for customizing the website in addition to those of fundamental, information-providing, and service-delivering facilities studied here.

## APPENDIX

- Appendix A. Modified checklist for evaluating the facilities of the commercial websites of Iranian top 100 companies

- Appendix B. Association Rules (sorted by the rate of confidence)

Appendix A. Modified checklist for evaluating the facilities of the commercial websites of Iranian top 100 companies

| Section 1) basic and technical-related facilities | | |
|---|---|---|
| 1- Does the website of the company have a reasonable load-time? (less than 30 seconds on a 56kbps internet connection) | Y | N |
| 2- Does the website of the company have a search-inside facility? | Y | N |
| 3- Does the website of the company have detailed submenus for easier access to different parts? | Y | N |
| 4- Does the website of the company have a site-map for easier navigation through it? | Y | N |
| 5- Does the website of the company have specific page-titles rather than codes or numbers for each page? | Y | N |
| **Section 2) Information-providing facilities** | | |
| 6- Does the website of the company have an "ABOUT US" section? | Y | N |
| 7- Does the website of the company give a general introduction of its field? | Y | N |
| 8- Does the website of the company have detailed description of its products/services? | Y | N |
| 9- Does the website of the company have a separate part as research department? | Y | N |
| 10- Does the website of the company have a "company in a glance" section? | Y | N |
| 11- Does the website of the company have an English page of company in a glance"? | Y | N |
| 12- Does the website of the company have an English homepage? | Y | N |
| 13- Does the website of the company have selected contents in English? | Y | N |
| 14- Does the website of the company have the contact info and introduction of the company's branches? | Y | N |
| 15- Does the website of the company have a part for advertising its own products/services? | Y | N |
| **Section 3) Service-delivering facilities** | | |
| 16- Does the website of the company have a CONTACT INFO page? | Y | N |
| 17 Does the website of the company have a Related Links section? | Y | N |
| 18- Does the website of the company have a Link to Others section? | Y | N |
| 19- Does the website of the company have links to news and related information? | Y | N |
| 20- Does the website of the company have a personnel log-in facility? | Y | N |

Appendix B. Association Rules (sorted by the rate of confidence)

| Rule number | Rule Antecedent | Rule Consequent | Confidence % | Support % |
|---|---|---|---|---|
| 1 | If the co. age is below 10 yrs, | Then its website will have "About-us" section. | 100 | 12.08 |
| 2 | If the co. age is below 10 yrs, | Then its website will have "Contact-us" section. | 100 | 12.08 |
| 3 | If the co. age is above 30 yrs, | Then its website will "Contact-us" section. | 100 | 49.45 |
| 4 | If the co. ownership is semi-private, | Then its website will have "Co. at a Glance" section. | 100 | 15.38 |
| 5 | If the co. ownership is semi-private, | Then its website will have "About-us" section. | 100 | 15.38 |
| 6 | If the co. ownership is private, | Then its website will have "Contact-us" section. | 100 | 30.76 |
| 7 | If the co. industry is products, | Then its website will have "Contact-us" section. | 100 | 48.35 |
| 8 | If the co. age is above 30 yrs AND the ownership is private, | Then its website will have "Contact-us" section. | 100 | 12.08 |
| 9 | If the co. age is 11-29 yrs AND the ownership is governmental, | Then its website will have "brief interdiction of co.'s field". | 100 | 21.97 |
| 10 | If the co. age is 11-29 yrs AND the ownership is governmental, | Then its website will have "About-us" section. | 100 | 21.97 |
| 11 | If the co. age is above 30 yrs AND the ownership is governmental, | Then its website will have "Contact-us" section. | 100 | 30.76 |
| 12 | If the co. age is 11-29 yrs AND the industry is products, | Then its website will have "About-us" section. | 100 | 16.48 |
| 13 | If the co. age is 11-29 yrs AND the industry is products, | Then its website will have "Contact-us" section. | 100 | 16.48 |
| 14 | If the co. age is above 30 yrs AND the industry is products, | Then its website will have "Contact-us" section. | 100 | 28.57 |
| 15 | If the co. age is above 30 yrs AND the industry is services, | Then its website will have "Contact-us" section. | 100 | 20.87 |
| 16 | If the co. ownership is private AND the industry is services, | Then its website will have "Contact-us" section. | 100 | 20.87 |
| 17 | If the co. ownership is governmental AND the industry is products, | Then its website will have "Contact-us" section. | 100 | 32.96 |
| 18 | If the co. ownership is governmental AND the industry is services, | Then its website will have reasonable load-time. | 100 | 20.87 |

| 19 | If the co. ownership is governmental AND the industry is services, | Then its website will have "brief interdiction of co.'s field". | 100 | 20.87 |
|---|---|---|---|---|
| 20 | If the co. ownership is governmental AND the industry is services, | Then its website will have "About-us" section. | 100 | 20.87 |
| 21 | If the co. ownership is governmental, | Then its website will have "About-us" section. | 97.95 | 53.84 |
| 22 | If the co. industry is products, | Then its website will have "About-us" section. | 97.72 | 48.35 |
| 23 | If the co. age is 11-29 yrs, | Then its website will have "About-us" section. | 97.14 | 38.46 |
| 24 | If the co. ownership is governmental AND the industry is products, | Then its website will have "About-us" section. | 96.66 | 32.96 |
| 25 | If the co. age is above 30 yrs AND the ownership is governmental, | Then its website will have reasonable load-time. | 96.42 | 30.76 |
| 26 | If the co. age is above 30 yrs AND the ownership is governmental, | Then its website will have "Abour-us" section. | 96.42 | 30.76 |
| 27 | If the co. age is above 30 yrs AND the industry is products, | Then its website will have "Co. at a Glance" section. | 96.15 | 28.57 |
| 28 | If the co. age is above 30 yrs AND the industry is products, | Then its website will have "About-us" section. | 96.15 | 28.57 |
| 29 | If the co. ownership is governmental, | Then its website will have "brief introduction of co.'s field". | 95.91 | 53.84 |
| 30 | If the co. ownership is governmental, | Then its website will have "Contact-us" section. | 95.91 | 53.84 |
| 31 | If the co. industry is services, | Then its website will have "Contact-us" section. | 95.74 | 51.64 |
| 32 | If the co. age is 11-29 yrs AND the industry is services, | Then its website will have "Co. at a Glance" section. | 95 | 21.97 |
| 33 | If the co. age is 11-29 yrs AND the industry is services, | Then its website will have "About-us" section. | 95 | 21.97 |
| 34 | If the co. age is above 30 yrs AND the industry is services, | Then its website will have "English homepage". | 94.73 | 20.87 |
| 35 | If the co. age is above 30 yrs AND the industry is services, | Then its website will have "brief interdiction of co.'s field". | 94.73 | 20.87 |
| 36 | If the co. ownership is governmental AND the industry is services, | Then its website will have "selected contents in English". | 94.73 | 20.87 |
| 37 | If the co. ownership is governmental AND the industry is services, | Then its website will have "English homepage". | 94.73 | 20.87 |

| 38 | If the co. age is 11-29 yrs, | Then its website will have "Contact-us" section. | 94.28 | 38.46 |
|---|---|---|---|---|
| 39 | If the co. industry is services, | Then its website will have "About-us" section. | 93.61 | 51.64 |
| 40 | If the co. age is above 30 yrs, | Then its website will have "brief interdiction of co.'s field". | 93.33 | 49.45 |
| 41 | If the co. age is above 30 yrs, | Then its website will have "About-us" section. | 93.33 | 49.45 |
| 42 | If the co. age is 11-29 yrs AND the industry is products, | Then its website will have "brief interdiction of co.'s field". | 93.33 | 16.48 |
| 43 | If the co. ownership is governmental AND the industry is products, | Then its website will have "brief interdiction of co.'s field". | 93.33 | 32.96 |
| 44 | If the co. industry is products, | Then its website will have "brief interdiction of co.'s field". | 93.18 | 48.35 |
| 45 | If the co. ownership is semi-private, | Then its website will have "research department" section. | 92.85 | 15.38 |
| 46 | If the co. ownership is semi-private, | Then its website will have "related links" section. | 92.85 | 15.38 |
| 47 | If the co. ownership is semi-private, | Then its website will have detailed submenus. | 92.85 | 15.38 |
| 48 | If the co. age is above 30 yrs and the ownership is governmental, | Then its website will have "brief interdiction of co.'s field". | 92.85 | 30.76 |
| 49 | If the co. age is above 30 yrs and the ownership is governmental, | Then its website will have "Co. at a Glance" section. | 92.85 | 30.76 |
| 50 | If the co. age is above 30 yrs and the industry is products, | Then its website will have "brief interdiction of co.'s field". | 92.30 | 28.57 |
| 51 | If the co. ownership is governmental, | Then its website will have reasonable load-time. | 91.83 | 53.84 |
| 52 | If the co. industry is services, | Then its website will have "Co. at a Glance" section. | 91.48 | 51.64 |
| 53 | If the co. age is 11-29 yrs, | Then its website will have "brief interdiction of co.'s field". | 91.42 | 38.46 |
| 54 | If the co. age is above 30 yrs, | Then its website will have "Co. at a Glance" section. | 91.11 | 49.45 |
| 55 | If the co. age is below 10 yrs, | Then its website will have "related links" section. | 90.90 | 12.08 |
| 56 | If the co. age is below 10 yrs, | Then its website will have "Co. at a Glance" section. | 90.90 | 12.08 |
| 57 | If the co. age is above 30 yrs AND the ownership is private. | Then its website will have specific page-titles. | 90.90 | 12.08 |

| 58 | If the co. age is above 30 yrs AND the ownership is private. | Then its website will have detailed submenus. | 90.90 | 12.08 |
|---|---|---|---|---|
| 59 | If the co. age is above 30 yrs AND the ownership is private. | Then its website will have "brief interdiction of co.'s field". | 90.90 | 12.08 |
| 60 | If the co. age is 11-29 yrs AND the ownership is governmental. | Then its website will have "selected contents in English". | 90 | 21.97 |
| 61 | If the co. age is 11-29 yrs AND the ownership is governmental. | Then its website will have "English homepage". | 90 | 21.97 |
| 62 | If the co. age is 11-29 yrs AND the ownership is governmental. | Then its website will have "Contact-us" section. | 90 | 21.97 |
| 63 | If the co. age is 11-29 yrs AND the industry is services. | Then its website will have "selected contents in English". | 90 | 21.97 |
| 64 | If the co. age is 11-29 yrs AND the industry is services | Then its website will have "English homepage". | 90 | 21.97 |
| 65 | If the co. age is 11-29 yrs AND the industry is services | Then its website will have "related links" section. | 90 | 21.97 |
| 66 | If the co. age is 11-29 yrs AND the industry is services | Then its website will have reasonable load-time. | 90 | 21.97 |
| 67 | If the co. age is 11-29 yrs AND the industry is services | Then its website will have "brief interdiction of co.'s field". | 90 | 21.97 |
| 68 | If the co. age is 11-29 yrs AND the industry is services | Then its website will have "Contact-us" section. | 90 | 21.97 |